\documentclass[letterpaper, 10 pt,conference]{ieeeconf}
\IEEEoverridecommandlockouts                              

\usepackage{cite}
\usepackage{amsmath,amssymb,amsfonts}
\usepackage{graphicx}
\usepackage{textcomp}
\usepackage{xcolor}
\usepackage{algorithm}
\usepackage{algcompatible}
\usepackage{microtype}
\usepackage{array}
\usepackage{tikz,}
\usepackage{graphicx}
\usepackage[utf8]{inputenc}
\usetikzlibrary{positioning}
\usetikzlibrary{arrows}
\IEEEoverridecommandlockouts   
\usepackage{caption}
\usepackage{subcaption}
\newcolumntype{P}[1]{>{\centering\arraybackslash}p{#1}}
\newcolumntype{M}[1]{>{\centering\arraybackslash}m{#1}}

\usepackage{tabularx, booktabs}
\usepackage{multirow}
\newcolumntype{Y}{>{\centering\arraybackslash}X}
\usepackage{threeparttable}
\usepackage{hyperref}
\makeatletter
\newcommand{\removelatexerror}{\let\@latex@error\@gobble}
\makeatother
\usepackage{pifont}
\newcommand{\cmark}{\ding{51}}%
\newcommand{\xmark}{\ding{55}}%

\begin{document}

\title{\Large \bf
Reconstructing Transit Vehicle Trajectory Using High-Resolution GPS Data \thanks{\footnotesize \noindent \textcopyright 2023 IEEE. Personal use of this material is permitted. Permission from IEEE must be obtained for all other uses, in any current or future media, including reprinting/republishing this material for advertising or promotional purposes, creating new collective works, for resale or redistribution to servers or lists, or reuse of any copyrighted component of this work in other works.}
}

\author{Yuzhu Huang$^{1}$, Awad Abdelhalim$^{2}$, Anson Stewart$^{2}$, Jinhua Zhao$^{2}$, Haris Koutsopoulos$^{3}$  
\thanks{$^{1}$ Department of Civil and Environmental Engineering, Massachusetts Institute of Technology {\tt\small yuzhuh@mit.edu}}%
\thanks{$^{2}$ Department of Urban Studies and Planning, Massachusetts Institute of Technology {\tt\small awadt@mit.edu}; {\tt\small ansons@mit.edu}; {\tt\small
jinhua@mit.edu}}%
\thanks{$^{2}$ Department of Civil and Environmental Engineering, Northeastern University {\tt\small h.koutsopoulos@northeastern.edu}}%
}

\maketitle



\begin{abstract}
High-resolution location (``heartbeat'') data of transit fleet vehicles is a relatively new data source for many transit agencies. On its surface, the heartbeat data can provide a wealth of information about all operational details of a recorded transit vehicle trip, from its location trajectory to its speed and acceleration profiles. Previous studies have mainly focused on decomposing the total trip travel time into different components by vehicle state and then extracting measures of delays to draw conclusions on the performance of a transit route. This study delves into the task of reconstructing a complete, continuous, and smooth transit vehicle trajectory from the heartbeat data that allows for the extraction of operational information of a bus at any point in time into its trip. Using only the latitude, longitude, and timestamp fields of the heartbeat data, the authors demonstrate that a continuous, smooth, and monotonic vehicle trajectory can be reconstructed using local regression in combination with monotonic cubic spline interpolation. The resultant trajectory can be used to evaluate transit performance and identify locations of bus delays near infrastructure such as traffic signals, pedestrian crossings, and bus stops.
\end{abstract}

\section{Introduction}
\label{intro}
\subsection{Background}
As more sensors and devices are installed on transit vehicles for monitoring various aspects of transit fleet operations, higher quality and more granular data become available to transit agencies. One recent data source addition is the high-resolution GPS data of vehicle location, often called ``second-by-second data" (referred to as ``heartbeat data" hereinafter), that records the historical location and other metadata of each transit vehicle at almost every second. 


While analysts can use operational metrics obtained from analyzing stop-level Automatic Vehicle Location (AVL) data to identify areas with poor performance, it is difficult to determine what exactly contributes to the poor performance in each area. As an example, an analyst may find that the average speed of buses is particularly low within one stop-to-stop segment, but the low average speed could be caused by slow-moving traffic due to congestion, or by stopping delays incurred by traffic signals. The average speed information alone is not enough to determine which one of the two is the primary source of delay, thus making it challenging to pinpoint the most effective traffic intervention or transit improvement strategy.

In comparison, the heartbeat data offers timestamps of vehicles not only at bus stops, but also at locations in between them. Therefore, the heartbeat data contains a richer set of data than AVL and would allow analysts to understand the exact behaviors of vehicles at any location along its route. Such detailed information offers an opportunity to uncover the interactions between buses and other road infrastructure and points of conflict besides bus stops that could impede bus movement such as traffic signals, pedestrian crossings, etc.

\subsection{Motivation}
At first glance, the heartbeat data seems to provide information about the exact location of each vehicle in such detail that it should be able to tell everything an analyst would wish to know about a vehicle's trip. However, closer examination of the data would show that the coordinate data can be noisy and vehicles may appear to straddle the road network, and the data recorded can have inconsistent frequency, making it difficult to directly read the exact location and speed of the vehicle from the heartbeat data at every point of interest.

The task of extracting operational information from the heartbeat data would be easier if a complete transit vehicle trajectory could be reconstructed from raw heartbeat data. Such a trajectory would provide reliable information about the location, speed, and acceleration of the transit vehicle at any time and distance into the trip, thus offering valuable information to analysts trying to understand how transit vehicles interact with the built environment at a very granular level. 

This research features the following contributions:
\begin{itemize}
    \item A concise definition for an ideal transit vehicle trajectory is proposed;
    \item A process for how noisy and intermittent heartbeat data can be converted to a series of discrete timestamped location data is demonstrated;
    \item Several smoothing algorithms that convert the discrete data to a continuous trajectory function are explored and evaluated.
\end{itemize}

To the authors' best knowledge, this research is the first of its kind that attempts to understand bus operations by reconstructing complete vehicle trajectories from heartbeat data.

\subsection{Related Work}

Researchers have explored various uses of transit heartbeat data to understand bus operations. Hall and Vyas calculated the average segment speeds of transit vehicles by dividing segment length by non-dwell travel times and evaluated how well transit vehicle speeds can represent the congestion state of general traffic \cite{hall2000buses}. Colghlan et al. calculated the average speed of the entire trip by dividing the total trip length by the total trip time and compared the result to an ideal speed to draw a conclusion about queueing delay  \cite{coghlan2019assigning}. Aemmer et al. used high-resolution GTFS-RT data to calculate vehicle speeds by taking the slope between two time-distance data points and used the difference between the said speed and free-flow speed to categorize vehicle delays \cite{aemmer2022measurement}. Lind and Reid calculated vehicle speeds similarly and made conclusions about local and rapid bus services by comparing the speed and reliability patterns \cite{lind2021diagnosing}. These researchers have relied on drawing conclusions about bus operations by calculating an average vehicle speed within a segment or between data records rather than attempting to reconstruct a complete speed profile of the vehicle. Cathey and Deiley constructed speed profiles of transit vehicles on freeways using the Kalman Filter and concluded that granular AVL data gathered from buses can serve as an additional data source for freeway performance monitoring, but did not address the issue of negative speeds resulting from the filtering process \cite{cathey2002transit}. 

As reviewed by Li et al., researchers in the non-transit space have proposed smoothing algorithms and imputation methods to reconstruct complete vehicle trajectories using vehicle location data of general-purpose vehicles \cite{LI2020225}. Toledo et al. proposed the use of local regression as the smoothing algorithm to minimize the measurement errors of vehicle location data and to construct a smooth and monotonic high-order polynomial as the vehicle trajectory\cite{ToledoTomer2007EoVT}. Venthuruthiyil et al. further discussed the strategies for selecting the optimal window size and polynomial order in reconstructing the trajectory function \cite{Venthuruthiyil}. Although the researchers demonstrated the applicability of local regression in retrieving complete vehicle trajectories, they did not address the lack of guarantee for the differentiability of high-order polynomials. Rafati Fard et al. presented a method for reconstructing the trajectories of vehicles captured in aerial photography using a two-step wavelet analysis technique that reduces measurement errors but did not discuss whether the reconstructed location trajectory is guaranteed to be non-decreasing \cite{RAFATIFARD2017150}. Wu et al. explored reconstructing traffic-related spatial-temporal datasets by imputing missing values with Graphical Neural Networks but presented applications only in reconstructing complete speed profiles rather than location profiles \cite{Wu_Zhuang_Labbe_Sun_2021}. 

Therefore, this research attempts to address the gap in the literature by presenting methods to reconstruct a continuous, smooth, and monotonic transit vehicle trajectory directly from high-resolution GPS coordinate data to enable the extract of the location, speed, and acceleration of a bus at any point in time into its trip.

\section{From Raw Coordinates to Time-Distance Data}
\subsection{Notation and Formulation}
The goal in reconstructing the vehicle trajectory using heartbeat data is to convert the series of timestamped coordinates into a continuous function that maps every point in time into a bus trip, i.e. ``time into trip", to a ``distance into trip" value. With classic vehicle kinematics equations, the time-to-distance mapping offered by the trajectory function will then allow for the derivation of the vehicle speed and acceleration profiles.

Suppose the heartbeat data of a bus trip contains a series of $n$ timestamped coordinates, i.e. 
\begin{equation}\label{eq:c_vector}
    C=[C_1, C_2, ..., C_n]^T
\end{equation}
 recorded at timestamps
 \begin{equation}\label{eq:S_vector}
     S = [S_1, S_2, ..., S_n]^T,
 \end{equation}
where for each $i \in \{1, 2, ..., n\}$, $C_i$ is a pair of latitude and longitude values $(lat_i, lon_i)$ and timestamp $S_i$ is the date and time at which the location of the bus is recorded as coordinate $C_i$ by the onboard device.

Through a map-matching process, the series of raw coordinates, $C$, is converted to a series of map-matched coordinates, $M$. Using information about the characteristics of the road network, the series of timestamps $S$ and the map-matched coordinates $M$ can be converted to time into trip values
\begin{equation}\label{eq:ti_vector}
    T=[t_1, t_2, ..., t_n]^T,
\end{equation}
and distance into trip values
\begin{equation}\label{eq:di_vector}
    D=[d_1, d_2, ..., d_n]^T.
\end{equation}
Detailed information about how the map-matching process can be carried out is described in Section \ref{sec:map_matching}. 

The series of discrete time-distance data can be used to infer the location of the vehicle at time points in the complement set $T^C = \{t\in T_R: t\notin T\}$, where $T_R$ is the set of real values within the time range $[0, t_n]$. To do so, an interpolation method or smoothing method is needed in order to impute the distance into trip values at time points in $T^C$. The algorithm takes $T$ and $D$ as input, and depending on the assumption around the error term associated with each value in $D$, outputs a function $f: T_R \to X$, where $X$ is a real set of true distance into trip values corresponding to each time into trip value in $T_R$.

The continuous function $f$ provides a mapping from time $T_R$ to distance $X$, and can be alternatively expressed as
\begin{equation}\label{eq:xt}
    x(t) = f(t).
\end{equation}
The speed profile of the vehicle can then be derived as
\begin{equation}\label{eq:vt}
    v(t) = \frac{d}{dt}x(t) = f'(t),
\end{equation}
and the acceleration profile is
\begin{equation}\label{eq:at}
    a(t) = \frac{d^2}{dt^2}x(t) = f''(t).
\end{equation}

\begin{figure}[h]
    \centering
    \includegraphics[width=0.80\linewidth]{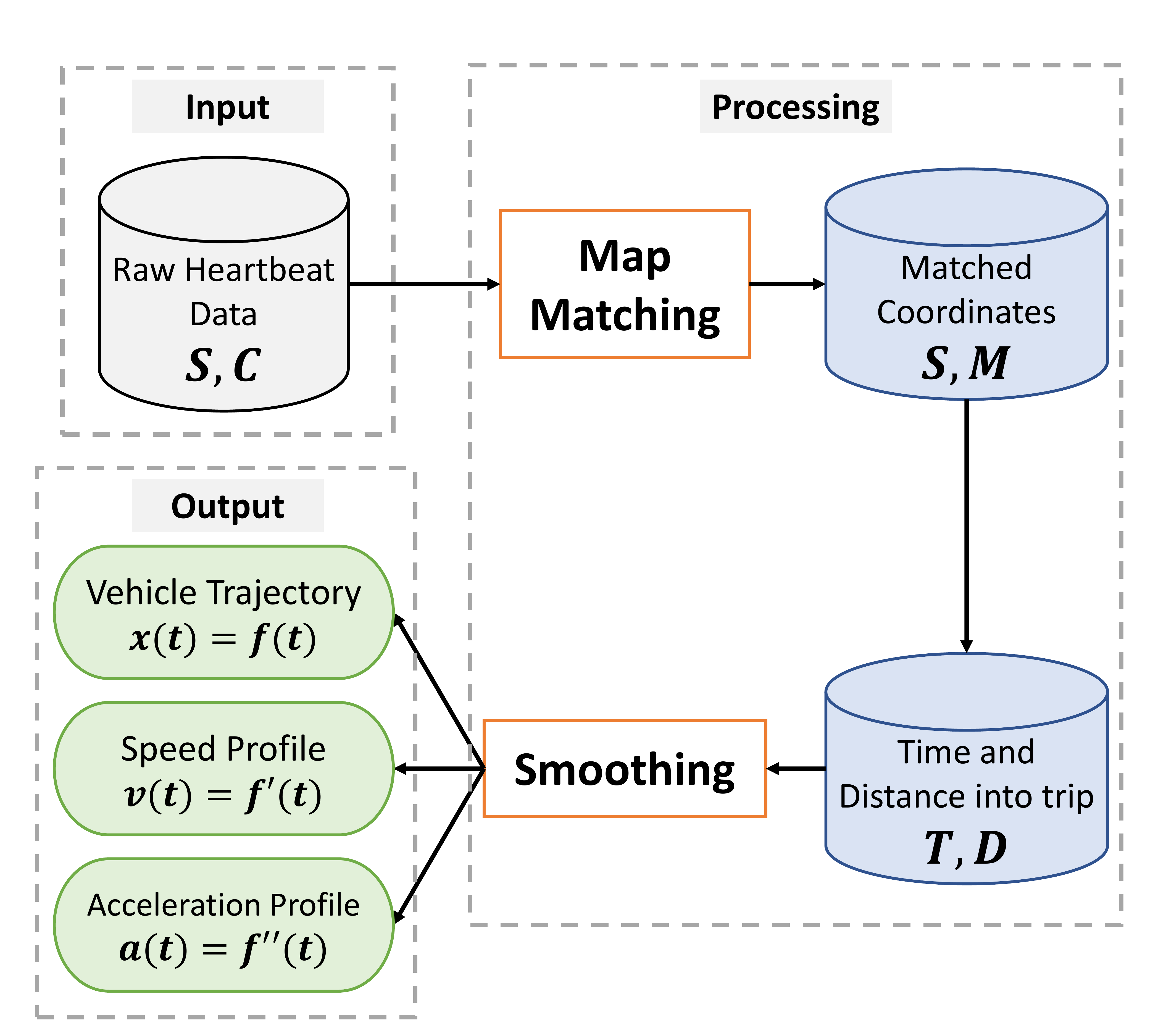}
    \caption{Flow chart of data processing.}
    \label{fig:flow_chart} 
\end{figure}

The entire process of reconstructing a complete vehicle trajectory from timestamped raw coordinates given by the heartbeat data is illustrated in the flow chart in Figure \ref{fig:flow_chart}.

\subsection{Map Matching}\label{sec:map_matching}
The raw coordinates recorded in the heartbeat data sometimes land in areas beyond the road network rather than within actual road segments. To mitigate the issue, a map-matching process is needed to ``snap" each coordinate point to the closest road segment to restore the location information as much as possible. 

The map-matching tool used by the authors is the Valhalla map engine, which offers the ``trace attributes" service that allows for the conversion of raw coordinates $C$ to map-matched coordinates $M$ \cite{valhalla, osm}. The service also returns the ID of the specific road segment, $r_i$, along the inferred path that each matched coordinate $M_i$ lies on, as well as the most probable normalized position, $p_i$, along the road segment $r_i$. The normalized position value $p_i$ is given so that $0\le p_i\le 1$. $p_i=0$ means the coordinate $M_i$ is at the beginning of the segment $r_i$, while $p_i=1$ means the end. 

For each bus trip, the vehicle is expected to run on a predefined and fixed route, so the ordered list of possible road segments that make up the pattern of the bus trip is known and denoted by 
\begin{equation}\label{eq:R_vector}
    \mathbf{R} = \langle R_1, R_2, ..., R_s\rangle,
\end{equation}
where $s$ is the total number of road segments in the list, and segment $R_b$ is downstream of $R_a$ for $b>a$. The map matched coordinate $M_i$ is considered valid only if $r_i\in \mathbf{R}$. 


When analyzing the heartbeat data of a single bus trip, the trip is recorded to start at timestamp $S_1$ and ends at timestamp $S_n$. Therefore, each timestamp $S_i$ can be converted to a ``time into trip" value $t_i$ through the simple calculation:
\begin{equation}\label{eq:ti}
    t_i = S_i - S_1.
\end{equation}

For every road segment in $\mathbf{R}$, its features including length are available from OpenStreetMap \cite{osm}. With this information, each matched coordinate of the bus can be converted to a distance value relative to the starting location of the trip.  Denote the length of segment $R_j$ by $L_j$ and the index of the road segment $r_i$ in the set $\mathbf{R}$ by $id_i$ (i.e. $r_i = R_{id_i}$), then a ``distance into trip" value can be calculated as follows:
\begin{equation}\label{eq:di}
    d_i = \sum\limits_{j=1}^{id_i-1} L_{j} + L_{id_i} \times p_i - L_{id_1} \times p_1
\end{equation}
Note that the last term in Equation \ref{eq:di} is needed to ensure that $d_1=0$ and that every distance into trip value is measured with respect to the first data point.

Combining the time into trip and distance into trip information calculated using Equations \ref{eq:ti} and \ref{eq:di}, a discrete series of time-distance data can be obtained from the map-matched coordinates.


\subsection{Properties of An Ideal Bus Trajectory}
The goal of data smoothing is to obtain a trajectory that resembles the real-world behavior of the bus as realistically as possible. Therefore, a few properties must be considered for the result to be representative of an actual transit vehicle trip. These properties include:
\begin{itemize}
    \item 
    The trajectory should be non-decreasing, i.e. the distance into trip at a later time point should not be smaller than the distance into trip at a previous time point.
    \item
    The trajectory is made up of composite cubic polynomials. The position of the vehicle is a result of control inputs and follows vehicle kinematics, and as Nagy et al. reported, cubic polynomials are the lowest order curves that can be used to generate the trajectory of car-like robots \cite{nagy}.
    \item
    The trajectory function should be continuous and at least once differentiable so that the speed profiles are also continuous and can be easily retrieved from the function.
\end{itemize}

Since a suitable smoothing algorithm is the backbone of the process of constructing a continuous trajectory function from discrete time-distance data points, discussions around the exploration and evaluation of the algorithms are carried out in the following section separately.

\section{Trajectory Smoothing}\label{sec:traj_smoothing}

\subsection{Linear Interpolation}
The simplest method to construct a continuous curve on the time-space diagram using the discrete time-distance data obtained from the previous steps is through linear interpolation, or equivalently, connecting adjacent data points using line segments (LSEG).



An example snippet of the trajectory obtained from linear interpolation is shown in Figure \ref{fig:matched_traj_zoomedin_itsc}. Considering the three properties of an ideal bus trajectory introduced above, it can be noticed that the trajectory is continuous and monotonic but not smooth or differentiable. An approximated speed profile can be obtained, however, through forward differentiation of the time-distance function. An acceleration profile can then be similarly produced from the speed profile.




\subsection{Polynomial Cubic Interpolation}

To address the non-differentiability issue of linear interpolation, the Piecewise Cubic Hermite Interpolant (PCHIP) algorithm developed by Fritsch et al. is explored \cite{Fritsch1}. The PCHIP algorithm connects adjacent data points in $T$ and $D$ with a cubic polynomial function and enforces a smooth trajectory by ensuring that the first derivative is equal for the connecting polynomial functions on each side of a data point \cite{Fritsch2}. This maintains the cubic polynomial and monotonic properties of the trajectory.

Figure \ref{fig:pchip_traj_zoomedin_itsc} shows an example of the trajectory obtained using PCHIP. While the PCHIP algorithm guarantees a continuous first derivative, it does not ensure a continuous second derivative, resulting in a continuous speed profile but not a continuous acceleration profile.


\subsection{Local Regression}
Although the PCHIP algorithm allows for the construction of a trajectory that satisfies all three properties of an ideal trajectory, one of the fundamental assumptions that have to be made is that all the discrete distance into trip data points that serve as the input to the algorithm are the true distances and that there is no error associated with each distance value. Such an assumption may not be valid, however, if the measurement error of the onboard device is taken into consideration. 

A local regression process (LOCREG) can be used to estimate the true distance into trip of the bus vehicle at each recorded timestamp by weighing the importance of nearby data points based on their distance from the timestamp of interest, consequently smoothing out the transit vehicle trajectory \cite{cleveland}. 

Suppose the true distance of the vehicle at each time point $t_i$ is $x_i$ such that
\begin{equation}\label{eq:xi}
    x_i = d_i + \varepsilon_i,
\end{equation}
where $d_i$ is the measured distance at time $t_i$, and $\varepsilon_i$ is the measurement error for $d_i$. The objective of the local regression algorithm is to find a function $f:t\to x$ such that it solves the following minimization problem at each data point $(t_i, d_i)$:
\begin{equation}
    \min \sum_{j=1}^n w_{i, j} (d_i - f(t_i))^2,
\end{equation}
where the value of each weight term $w_{i, j}$ is determined by the selected bandwidth and kernel function. In this study, cubic polynomials, the tricube kernel proposed by Cleveland and a bandwidth of 20 data points are selected empirically in the estimation of each true distance $x_i$ \cite{cleveland}.

An example snippet of the trajectory obtained from local regression is shown in Figure \ref{fig:locreg_traj_zoomedin_itsc}. LOCREG is superior to LSEG and PCHIP in that it produces a continuous trajectory that takes into account the measurement error of each measured distance, but it does not necessarily guarantee differentiability or monotonicity of the overall function. This can result in negative speed values when taking the first derivative of the LOCREG trajectory and also presents challenges in deriving speed and acceleration profiles.


\subsection{Interpolation and Regression}
To address the issue with monotonicity from using local regression, an exploratory algorithm is experimented with. This algorithm, named LOCREG-PCHIP, combines the desired properties of both PCHIP and LOCREG, removes the non-monotonic sections of the LOCREG data, and fills in the now missing intervals by passing the remaining data through a monotonic interpolation algorithm such as PCHIP.

A complete workflow that starts from data smoothing and ends with filling in data to obtain a smooth and monotonic trajectory is detailed in Algorithm \ref{alg:locreg_pchip}. 
\begin{algorithm}
\caption{LOCREG-PCHIP}
\label{alg:locreg_pchip}
\hspace*{\algorithmicindent} Input: an $1\times n$ vector $T$ of time into trip values $t_1 < t_2 < ... < t_n$, an $1\times n$ vector $D$ of distance into trip values $d_1 \leq d_2 \leq ... \leq d_n$ for $i=1,..., n$ and $n>2$. 

\begin{algorithmic}[1]
\small
\STATE X = []
\STATE $f_{locreg}= LOCREG(T, D)$
\FOR{i = 1, 2, ..., n}
    \STATE $x_i = f_{locreg}(t_i)$
    \IF{$i > 1$ and $x_i < x_{i-1}$}
        \STATE $x_i = x_{i-1}$
    \ENDIF
    \STATE $X.append(x_i)$
\ENDFOR
\STATE $f=PCHIP(T,X)$
\STATE return $f$

\end{algorithmic}
\end{algorithm}

\begin{figure}[!h]
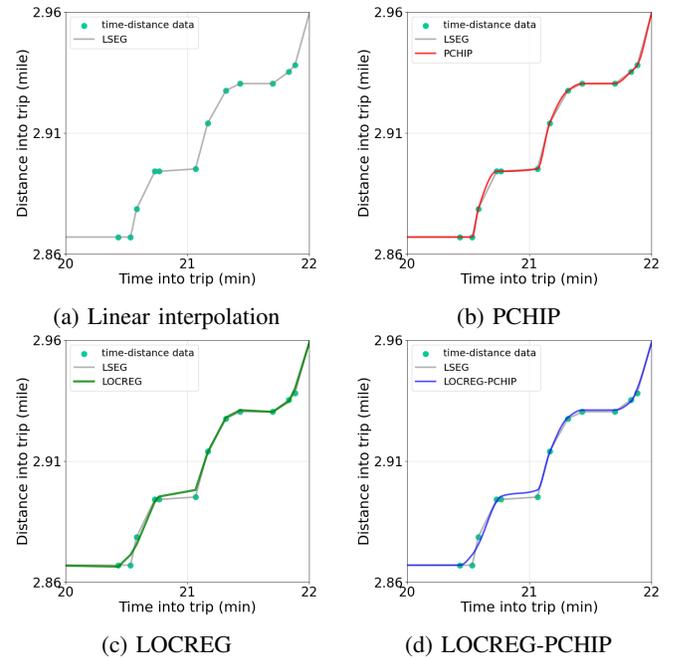

     \centering
     \begin{subfigure}[b]{0.23\textwidth}
         \centering
         \includegraphics[width=\linewidth]{fig/matched_traj_zoomedin_itsc.pdf}
         \caption{Linear interpolation}
         \label{fig:matched_traj_zoomedin_itsc}
     \end{subfigure}
     \hfill
     \begin{subfigure}[b]{0.23\textwidth}
         \centering
         \includegraphics[width=\linewidth]{fig/pchip_traj_zoomedin_itsc.pdf}
         \caption{PCHIP}
         \label{fig:pchip_traj_zoomedin_itsc}
     \end{subfigure}\\
     
    \begin{subfigure}[b]{0.23\textwidth}
         \centering
         \includegraphics[width=\linewidth]{fig/locreg_traj_zoomedin_itsc.pdf}
         \caption{LOCREG}
         \label{fig:locreg_traj_zoomedin_itsc}
     \end{subfigure}
     \hfill
     \begin{subfigure}[b]{0.23\textwidth}
         \centering
         \includegraphics[width=\linewidth]{fig/pchip_locreg_traj_zoomedin_itsc.pdf}
         \caption{LOCREG-PCHIP}
         \label{fig:pchip_locreg_traj_zoomedin_itsc}
     \end{subfigure}
    \caption{Trajectory constructed from different algorithms.}
    \label{fig:all_trajs}
\end{figure}

An example snippet of the trajectory obtained from local regression is shown in Figure \ref{fig:pchip_locreg_traj_zoomedin_itsc}. The trajectory obtained using LOCREG-PCHIP preserves the true distance values estimated from local regression at all time points except for those where the monotonicity principle is violated, in which case the distance value is chosen to be equal to the largest distance value observed prior to the said time points. Since the algorithm ends with constructing a PCHIP function using the modified distance values, the resultant trajectory function is both monotonic and differentiable.

\section{Evaluation and Validation}
The ground truth location, speed and acceleration data of a bus is not often available. The following sections offer some alternative validation strategies. The discussion uses a sample trajectory that is considered representative of a typical bus trip as described in Section \ref{sec:data_source}, thus the conclusions are assumed to be applicable to other bus trips as well.

\subsection{Validation of the Speed Profile}
\begin{figure}[h]
    \centering
    \includegraphics[width=0.90\linewidth]{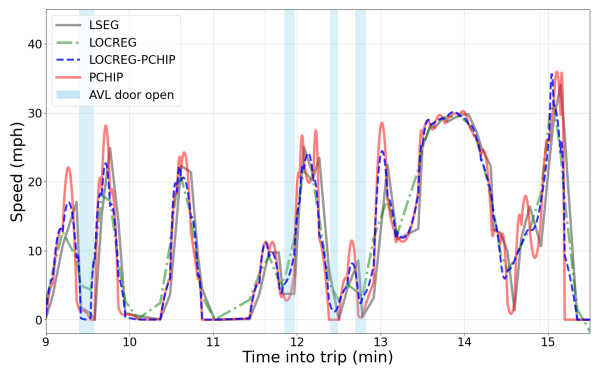}
    \caption{Comparison of all speed profiles.}
    \label{fig:all_speed_traj_zoomedin_thesis_itsc} 
\end{figure}

One way to validate the accuracy of the speed profile is to check if the speed from the calculated trajectory is indeed zero when the transit vehicle was recorded as having doors open at a bus stop in the AVL data. The speed profiles obtained from all aforementioned algorithms can be displayed over color bands of door-open intervals in the AVL data as illustrated in Figure \ref{fig:all_speed_traj_zoomedin_thesis_itsc}. 

In order to measure how well the speed trajectory aligns with the AVL stop events, the number of integer seconds within AVL door-open intervals that correspond to a zero speed on the speed profile can be calculated. The notion of ``zero speed" is loosely defined as a speed below which a vehicle is considered not traveling to account for the additional errors in the distance values that were not accounted for in the smoothing process. 

\begin{figure}[h]
    \centering
    \includegraphics[width=0.65\linewidth]{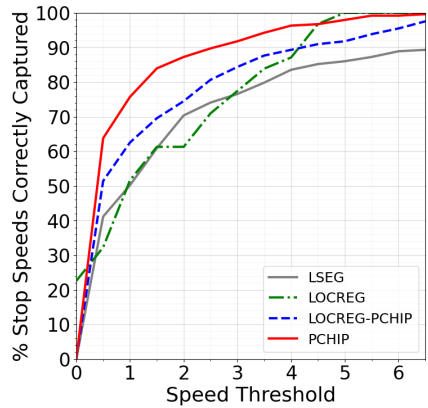}
    \caption{Percentage of AVL door-open instances at which speeds are correctly captured vs. the ``stop speed" threshold.}
    \label{fig:qc_speed_thesis} 
\end{figure}

As summarized in Table \ref{tab:qc_table}, if the ``stop speed" is defined as below 5 mph, the speed at over 90\% of the integer timestamps labeled as door-open in the AVL data are correctly captured using the LOCREG, LOCREG-PCHIP, and PCHIP algorithms. A more intuitive plot comparing the percentage of stop speed correctly captured by each algorithm is shown in Figure \ref{fig:qc_speed_thesis}.

\begin{table}[h]
\caption{Percentage of AVL door-open integer timestamps at which speeds are correctly captured by each algorithm.}
    \centering
    \small
    \begin{tabular}{M{0.3\linewidth}  M{0.1\linewidth} M{0.1\linewidth} M{0.1\linewidth}}
        \toprule
        \multirow{2}{*}{Algorithm} & \multicolumn{3}{c}{Stop-Speed Threshold (mph)}\\
        \cmidrule{2-4}
        {} & 0 & 3 & 5  \\
        \midrule
        LSEG & 0 & 77 & 86 \\
        LOCREG & 7 & 77 & 100\\
        LOCREG-PCHIP & 0 & 84 & 92\\
        PCHIP & 0 & 92 & 98\\
        \bottomrule
    \end{tabular}
    \label{tab:qc_table}
\end{table}

\subsection{Validation of the Acceleration Profile}
\begin{figure}[!h]
    \centering
    \includegraphics[width=0.85\linewidth]{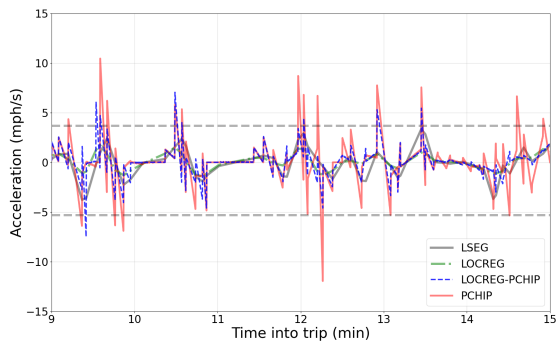}
    \caption{Comparison of all acceleration profiles.}
    \label{fig:all_acc_traj_zoomedin_itsc} 
\end{figure}

Another way to check the performance of the algorithm is to examine the percentage of accelerations that are beyond a reasonable threshold. Based on the research by Kirchner et al., the maximum bus acceleration is $0.17g=3.7$ miles-per-hour-per-second (mphps) and maximum deceleration is $-0.24g=-5.3$ mphps \cite{kirchner2014characterisation}. As indicated by the acceleration profiles shown in Figure \ref{fig:all_acc_traj_zoomedin_itsc}, LOCREG and LOCREG-PCHIP produce acceleration profiles that are more reasonable than the other two methods. This is verified by the percentage of unreasonable accelerations summarized in Table \ref{tab:acc_qc_table}, where the authors find that only 1.3\% of the accelerations from LOCREG-PCHIP and none from LOCREG are beyond the reasonable threshold, whereas 2.1\% and 5.3\% of accelerations from LSEG and PCHIP, respectively, are unreasonable. 

\begin{table}[h]
\caption{Percentage of trajectory data of which accelerations are beyond the $[-5.3, 3.7]$ mphps threshold.}
    \centering
    \small
    \begin{tabular}{M{0.3\linewidth}  M{0.4\linewidth}}
        \toprule
        Algorithm & \% unreasonable acceleration\\
        \midrule
        LSEG & 2.1 \\
        LOCREG & 0.0\\
        LOCREG-PCHIP & 1.3\\
        PCHIP & 5.3\\
        \bottomrule
    \end{tabular}
    \label{tab:acc_qc_table}
\end{table}

Given the discussion above, the trajectories produced by LOCREG, LOCREG-PCHIP, and PCHIP are better compared to those from LSEG when validated against stop-level AVL data and reasonable acceleration thresholds.

\subsection{Selection of the Best Algorithm}
The ``best algorithm" should produce trajectories that not only perform reasonably well in the evaluation of speed and acceleration profiles but also satisfy all three characteristics of an ideal trajectory. As shown in Table \ref{tab:alg_comparison}, the algorithms that satisfy these criteria are LOCREG-PCHIP and PCHIP.

\begin{table}[h]
\caption{Evaluation of algorithms.}
    \centering
    \small
    \begin{threeparttable}
    \begin{tabular}{M{0.16\linewidth}  M{0.06\linewidth} M{0.06\linewidth} M{0.06\linewidth} M{0.06\linewidth} M{0.06\linewidth} M{0.06\linewidth} M{0.06\linewidth}} 
        \toprule
        \multirow{2}{*}{Algorithm} & \multicolumn{7}{c}{Evaluation Criteria}\\
        \cmidrule{2-8}
        {} & MON\tnote{1} & CUB\tnote{2} & DIFF\tnote{3} & ERR\tnote{4} & AVL (\%)\tnote{5} & ACC (\%)\tnote{6} & Best\\
        \midrule
        LSEG & \cmark & \xmark & \xmark & \xmark & 86 & 98 \\
        LOCREG & \xmark & \xmark & \xmark & \cmark & 100 & 100 \\
        LOCREG-PCHIP & \cmark & \cmark & \cmark & \cmark & 92 & 99 & \cmark\\
        PCHIP & \cmark & \cmark & \cmark & \xmark & 98 & 95 & \\
        \bottomrule
    \end{tabular}
    \begin{tablenotes}\footnotesize
    \item[1] MON: the trajectory is non-decreasing
    \item[2] CUB: the trajectory is made up of cubic polynomials
    \item[3] DIFF: the trajectory is once differentiable
    \item[4] ERR: minimizes measurement error
    \item[5] AVL (\%): the percentage of AVL dwell activities correctly captured
    \item[6] ACC (\%): the percentage of accelerations within a reasonable threshold
    \end{tablenotes}
    \end{threeparttable}
    \label{tab:alg_comparison}
\end{table}

Comparing the LOCREG-PCHIP algorithm with PCHIP, LOCREG-PCHIP is preferable because it is able to predict the true distances from observed distances using the information provided by adjacent data points by taking advantage of the LOCREG algorithm, rather than trusting each observation a hundred percent. Therefore, it is decided that LOCREG-PCHIP is the best overall algorithm for reconstructing bus trajectories.

\section{Case Study}
To demonstrate how a complete bus trajectory can be reconstructed using heartbeat data, a case study using real-world data is provided in the following section. The case study aims at providing a concrete example of the data processing procedures discussed in the previous sections and offers insight into how analysts can utilize the complete bus trajectories constructed using methods provided in this research to draw insight into transit operations.

\subsection{Data Source}\label{sec:data_source}
The heartbeat data analyzed in this study are archived snapshots of the GTFS-RT data taken from the public-facing API made available by the Massachusetts Bay Transportation Authority (MBTA). Each heartbeat data point records a timestamped location of the transit vehicle during its trip. An examination of the heartbeat data recorded over 12 weekdays of outbound Route 1 operated by the MBTA reveals that most data is recorded in intervals shorter than 10 seconds with a median frequency of 6 seconds, mode of 3 seconds and mean of 9 seconds.


For the purpose of this case study, the heartbeat data of one outbound Route 1 trip operated on a weekday morning (Monday, April 25, 2022, at 8 AM) is analyzed. 

\subsection{Reconstructing Vehicle Trajectory}
The heartbeat data of the analyzed trip contains 328 timestamped raw coordinates, a subset of which is shown in the $S$ and $C$ columns in Table \ref{tab:map_matching_example}. The raw coordinates are passed into the map-matching engine Valhalla \cite{valhalla}, which returns the corresponding matched coordinates shown in column $M$, the ID of the road segment in column $r$, and the position of each matched coordinate along its segment in column $p$.

\begin{table}[h]
    \centering
    \caption{An example map-matched coordinate table.}
    \small
    \begin{tabularx}{0.48\textwidth}{ P{0.01\textwidth}P{0.09\textwidth}P{0.09\textwidth}P{0.09\textwidth}P{0.01\textwidth}P{0.02\textwidth}}
        \toprule
        $i$ & $S$ & $C$ & $M$ & $r$ & $p$  \\
        \midrule
        0 & 2022-04-25 08:24:45 & (42.372642, -71.119048) & (42.372660, -71.119108) &0 & 0.639\\
        1 &	2022-04-25 08:24:50 & (42.372365, -71.119241) & (42.372373, -71.119267) &0 & 0.940\\
        2 &	2022-04-25 08:24:57 & (42.372246, -71.119324) & (42.372252, -71.119319) &2 & 0.463\\
        ...\\
        \bottomrule
    \end{tabularx}
    \label{tab:map_matching_example}
\end{table}

\begin{figure}[h]
    \centering
    \includegraphics[width=0.95\linewidth]{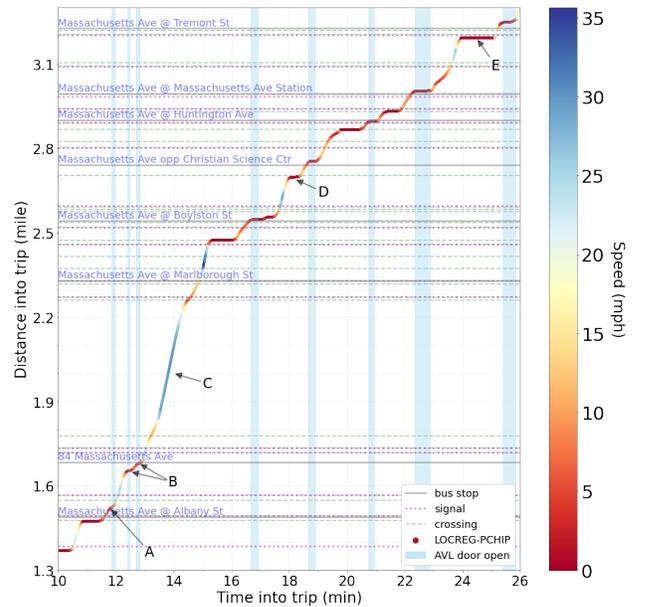}
    \caption{A snippet of a LOCREG-PCHIP trajectory used for the operational analysis of a sample weekday PM inbound trip of Route 1 operated by the MBTA.}
    \label{fig:detailed_traj} 
\end{figure}

Using Equations \ref{eq:ti} and \ref{eq:di}, the time into trip $T$ and distance into trip $D$ values of each data point can be calculated and passed to the LOCREG-PCHIP algorithm, a continuous vehicle trajectory color-coded by vehicle speed is obtained and shown in the time-space diagram in Figure \ref{fig:detailed_traj}. The location of road infrastructure such as bus stops, traffic signals, and pedestrian crossings can also be identified through the map matching process and plotted in the same figure.

The time-space diagram of a complete vehicle trajectory, in combination with AVL data, can provide a wealth of information regarding the bus operation and lend insight into the movement of the bus throughout its trip. 

\subsection{Qualitative Analysis of A Vehicle Trajectory}
Several observations of the operations of the sample bus trip are described below to showcase the information that a complete bus trajectory can provide.
\subsubsection{Stop Dwelling Activities}
As shown by the sections of the trajectory labeled $A$ in Figure \ref{fig:detailed_traj}, the trajectory captures the bus stopping at the far-side bus stop at Massachusetts Ave \& Albany St after it waits at the traffic signal just upstream of the bus stop.

The sections labeled $B$ show that the bus opened its door once upstream of the bus stop at 84 Massachusetts Avenue and once at the bus stop. Interestingly, this behavior could be due to the Massachusetts State regulation which requires bus drivers to stop and open the doors before a railroad track \cite{madpu}. Both of these stopping activities are validated by the door-open intervals recorded in the AVL data.
\subsubsection{Vehicle Speed}
The section labeled $C$ in Figure \ref{fig:detailed_traj} shows that the bus traveled at a speed of approximately 25-30 mph 1.9-2.2 miles into the trip (Harvard Bridge), faster than its speed at most other sections of the trip.
\subsubsection{Stopping at a Pedestrian Crossing}
From the section labeled $D$ in Figure \ref{fig:detailed_traj}, one can see that the vehicle stopped before a pedestrian crossing prior to stopping at the bus stop at Massachusetts Ave opposite the Christian Science Center.


\subsubsection{Stopping at A Traffic Signal}
The section labeled $E$ shows that the bus stopped at a traffic signal upstream of the far-side stop at Massachusetts Ave \& Tremont St. Besides clearly showing the existence of the stopping activity at the traffic signal, the trajectory also shows the duration that the bus stopped at the signal. Such information about how long buses stop at a signal can be valuable for the decision-making related to transit signal priority projects. 

\section{Conclusion and Discussion}
\label{sec:conclusion}
This study developed methodologies to reconstruct continuous, monotonic, and differentiable bus trajectories from noisy heartbeat data. The trajectory smoothing algorithm LOCREG-PCHIP was identified as the best algorithm that produces a trajectory satisfying ideal properties while performing well against expected speed and acceleration data. The continuous bus trajectories allow for the extraction of location, speed, and acceleration at any point in time into a trip.

Several limitations are present with the methodologies that could be addressed in future research. First, this study only explored using heartbeat data with an average frequency of below 10 seconds. Further research may be needed to determine how the algorithm will perform on data with larger time intervals. Secondly, none of the smoothing algorithms presented in this study guarantees that the trajectory is twice-differentiable, therefore the acceleration profile is not guaranteed to be smooth. Lastly, the validation method used in this study to evaluate the performance of the algorithm is merely checking whether the speed and acceleration values fall within acceptable thresholds. An ideal method, however, is to compare the location, speed, and acceleration of vehicles with ground-truth data collected from telematics devices.

Further research will explore how the analysis of multiple bus trajectories can be analyzed in batches to allow for the categorization and quantification of different types of delays encountered by buses of specific routes or corridors.

\bibliographystyle{IEEEtran}
\bibliography{itsc}
\end{document}